\documentclass{ws-rmp}

\newcommand{\rmi}{\mathrm{i}}
\newcommand{\RMI}{\mathrm{I}}
\newcommand{\RMJ}{\mathrm{J}}
\newcommand{\rmd}{\mathrm{d}}
\newcommand{\rme}{\mathrm{e}}
\newcommand{\A}{\mathbb{A}}
\newcommand{\B}{\mathbb{B}}

\begin{document}

\markboth{Jon Harrison}{Quantizing graphs, one way or two?}

%
\catchline{}{}{}{}{}
%

\title{Quantizing graphs, one way or two?}

\author{Jon Harrison}

\address{Department of Mathematics,
	Baylor University,
	Sid Richardson Building,
	1410 S.4th Street,
	Waco, TX 76706,
	USA\\
\email{jon\_harrison@baylor.edu}}

\maketitle

\begin{history}
\received{(Day Month Year)}
\revised{(Day Month Year)}
\end{history}

\begin{abstract}
Quantum graphs were introduced to model free electrons in organic molecules using a self-adjoint Hamiltonian on a  network of intervals.  A second graph quantization describes wave propagation on a graph by specifying scattering matrices at the vertices.  A question that is frequently raised is the extent to which these models are the same or complementary.  In particular, are all energy independent unitary vertex scattering matrices associated with a self-adjoint Hamiltonian?  Here we review results related to this issue.  In addition, we observe that a self-adjoint Dirac operator with four component spinors produces a secular equation for the graph spectrum that matches the secular equation associated with wave propagation on the graph when the Dirac operator describes particles with zero mass and the vertex conditions do not allow spin rotation at the vertices.    
\end{abstract}



\section{Introduction}\label{sec:intro}

Quantum graphs are a popular model of quantum mechanics in systems with complex geometry.  They date back to models of free electrons in organic molecules \cite{P36} and have a wide range of current applications in, for example, Anderson localization, quantum chaos, wave-guides, nanotechnology, and superconductors.  An excellent introduction is \cite {BerKuch}.  

In many applications two different approaches are employed to quantize a graph.  Firstly edges are associated with intervals of a fixed positive length defining a metric graph.  An operator, typically the negative Laplace, Schr\"odinger or Dirac operator, acts on functions defined on the collection of intervals with local vertex conditions matching values of the functions and their derivatives at the ends of the intervals where they meet, such that the operator is self-adjoint.  This was the model presented by Pauling in 1936 \cite{P36} and introduced in the mathematics literature in the 80's \cite{L80,R84}.  

An alternative approach to quantize a graph was identified by Chalker et. al. \cite{CC88,CS90} and independently in the context of studying the graph spectrum by Schanz and Smilansky \cite{SS00} and Tanner \cite{T01}. In this model waves propagate on the graph according to unitary scattering matrices which are specified at the vertices.  This second method of quantizing a graph is often convenient as one 
may be able 
to choose vertex scattering matrices in a way that simplifies the problem under investigation.  For example, a common choice of vertex scattering matrices are the discrete Fourier transform (DFT) matrices,
\begin{equation}
\sigma^{(v)} = \frac{1}{\sqrt{d}} \left( 
\begin{array}{ c c c c c }
1 & 1 & 1 & \dots & 1\\
1 & w & w^2 &\dots &w^{d-1} \\
1 & w^2 & w^4 &\dots &w^{2(d-1)} \\
\vdots & \vdots & \vdots & & \vdots \\
1 & w^{d-1} & w^{2(d-1)} &\dots &w^{(d-1)(d-1)} \\
\end{array}
\right)\ ,
\label{DFTmatrix}
\end{equation}
where $d$ is the degree of the vertex and $w=\exp (2\pi \rmi / d)$.  DFT scattering matrices have the advantage that the scattering is democratic, in the sense that the probability of scattering onto every edge is equal, $|\sigma^{(v)}_{ij}|^2=1/d$ for all $i,j$. 
Similarly, one might specify equi-transmitting scattering matrices at the vertices \cite{HSW07,KO16}.  Equi-transmitting matrices have the property that $|\sigma^{(v)}_{ii}|^2=0$ and $|\sigma^{(v)}_{ij}|^2=1/(d-1)$ for $i\neq j$.  Choosing such scattering matrices prevents back-scattering, providing a connection with the Ihara zeta function of the non-quantum graph \cite{Ter}, while incoming waves scatter to the other edges with equal probability.  


Importantly, however, as the scattering matrix reflects the physical nature of interactions at the vertices  it may not be possible to choose vertex scattering matrices freely in a particular graph model. 
For example, there may not even be a self-adjoint Laplacian with vertex matching conditions that produces scattering matrices with the desired features. 
So the two approaches to quantizing a graph, specifying a self-adjoint operator or specifying a set of unitary vertex scattering matrices, naturally leads to the question of the extent to which the approaches are equivalent or complimentary.\footnote{This issue was raised in the dynamical systems section at QMATH 15 where the discussion prompted this article.}  Discussion and results related to this issue have appeared in a number of places, for example \cite{B08,BerKuch,FKW07}.
  In this article we review these contributions before identifying a situation where the spectrum of a quantum graph Hamiltonian matches the spectrum of the graph quantized by specifying the vertex scattering matrices directly.  
  While similar constructions were presented in \cite{B08} the situation we describe is simple and a natural physical operator.    The desired correspondence appears for a Dirac operator with four component spinors describing particles with zero mass when the vertex conditions do not allow spin rotation at the vertices.  A Dirac operator with four component spinors is the natural version of a Dirac operator to consider if one thinks of restricting a Dirac operator in three dimensions to a one dimensional line, which keeps the physical interpretation of spin.  However, it is not the only construction of a Dirac operator on a graph as the Dirac algebra in one dimension has a faithful irreducible representation using two component spinors \cite{B08,BH03,C99,FKW07}.   

The article is organized as follows.  In Sec. \ref{sec:quantizations} we reprise the quantization of a graph with a self-adjoint operator and the wave-propagation quantization.  
In Sec. \ref{sec:lit} we review the relationship between these quantization schemes.  
Finally, in Sec. \ref{sec:Dirac construction} we describe the quantization with the Dirac operator where the
secular equation matches secular equations obtained from a wave-propagation quantization.  This demonstrates that the wave propagation quantization can be regarded as a subset of quantum graphs defined by self-adjoint Hamiltonians.

\section{Graph Quantizations}\label{sec:quantizations}

\begin{figure}
	\centerline{\includegraphics[width=1.75in]{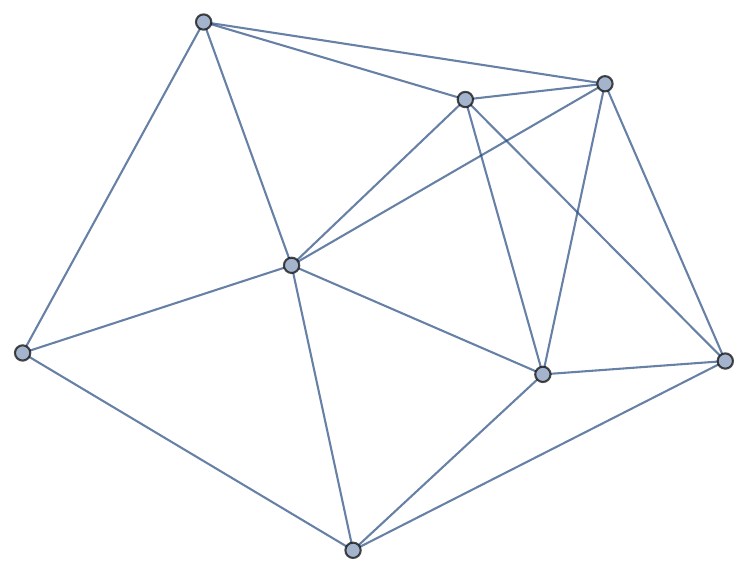}}
	\vspace*{8pt}
	\caption{A simple graph with $8$ vertices and $18$ edges.}\label{fig:graph}
\end{figure}

A \emph{graph} $G$ is a set of \emph{vertices} $\mathcal{V}=\{1, \dots , V \}$ and a set of \emph{edges} $\mathcal{E}$.  Edges join pairs of vertices so an edge $e=(u,v)$ is a pair of unordered vertices $u,v\in \mathcal{V}$, we say $u$ and $v$ are \emph{adjacent} and that $e$ is \emph{incident} with $u$ and $v$.  The degree of a vertex $d_v$ is the number of edges incident with $v$ and the total number of edges $|\mathcal{E}|=E$.   Graphs can have directed edges described by ordered pairs of vertices in which case they are digraphs.  However, our basic starting point will be to assume that $G$ is not a digraph, that $\mathcal{V}$ is finite, and that the graph is simple.  A \emph{simple} graph has no loops (edges do not join a vertex to itself) and pairs of vertices are connected by at most one edge.   If a graph is not simple it can be made simple by adding a vertex of degree two on any loops and multiple edges.    Fig. \ref{fig:graph} shows a simple graph.  Given such a graph $G$ we now describe two approaches used to obtain a quantum graph.

\subsection{A self-adjoint Hamiltonian on a metric graph}

The original setting for a quantum graph is to associate the edges with finite intervals
$[0,L_e]$ for $e\in \mathcal{E}$.  This turns $G$ into a \emph{metric graph} $\Gamma$ where the distance between points $x$ and $y$ on $\Gamma$ is the length of the shortest path joining $x$ and $y$.  We can envision the graph as a network of one-dimensional wires on which we want to model quantum mechanics.  When assigning the interval $[0,L_e]$ to the edge $e=(u,v)$ there is a choice for the direction of the coordinate $x_e\in [0,L_e]$. 
Ordinarily this choice does not affect the graph quantization.

The metric graph $\Gamma$ is quantized by defining a self-adjoint Hamiltonian operator, typically the Laplace, Schr\"odinger or Dirac operator.  For example, on each interval one considers the Laplace equation,
\begin{equation}\label{eq:Laplace eqn}
-\frac{\rmd^2 }{\rmd x_e^2} f_e(x_e) = k^2 f_e(x_e) \ ,
\end{equation}
with local vertex conditions at the ends of the interval that match values of the function and its derivatives meeting at a vertex.    At each vertex $v\in \mathcal{V}$ one defines vectors of values of the function at the ends of intervals incident at $v$ and the outgoing derivatives,
\begin{eqnarray}
	\mathbf{F}(v)&=&(f_{e_1}(0),\dots, f_{e_l}(0),f_{e_{l+1}}(L_{e_{l+1}}), \dots , f_{e_d} (L_{e_d})))^T \ , \\ 
	\mathbf{F'}(v)&=&(f'_{e_1}(0),\dots, f'_{e_l}(0),-f'_{e_{l+1}}(L_{e_{l+1}}), \dots , -f'_{e_d} (L_{e_d})))^T \ ,
\end{eqnarray}
where $v$ has degree $d$ and $v$ is located at $x_e=0$ on the first $l$ edges.  Then given a pair of $d\times d$ matrices $\A_v$ and $\B_v$ the vertex conditions are defined by the linear system,
\begin{equation}\label{eq:vertex condition}
\A_v \mathbf{F}(v) + \B_v \mathbf{F'}(v) = \mathbf{0} \ .
\end{equation}
The domain of the Hamiltonian is the subspace of $\bigoplus_{e\in \mathcal{E}} W^{2,2}([0,L_e])$ satisfy the vertex conditions, where $W^{2,2}$ is the Sobolev space of functions in $L^2$ along with their weak first and second derivatives.  The Laplacian is self adjoint if and only if  
$(\A_v,\B_v)$ has maximal rank and $\A_v\B_v^\dagger$ is Hermitian for all $v\in \mathcal{V}$ \cite {KS99}.

Self-adjoint realizations of the Laplace operator can be uniquely classified by choosing a $d_v\times d_v $ unitary matrix $U_v$ for each vertex \cite{BerKuch}  where one requires that the wavefunction satisfies vertex conditions,
\begin{equation}\label{eq:alt vertex conditions}
\rmi (U_v-\RMI ) \mathbf{F}(v) +(U_v +\RMI) \mathbf{F}'(v) =\mathbf{0} \ .
\end{equation}
Equivalently the vertex conditions of self-adjoint Laplacians can be described by three mutually orthogonal projectors onto Dirichlet, Neumann and Robin subspaces \cite{K04}.  

\subsection{Wave scattering quantum graphs}

To make the link with wave dynamics, solutions of the Laplace equation on an interval (\ref{eq:Laplace eqn}) are plane waves,
\begin{equation}
f_e(x_e) = a_e \rme^{\rmi k x_e} +a_{\bar{e}} \rme^{-\rmi k x_e} \ .
\end{equation} 
Defining vectors of coefficients of the incoming and outgoing plane waves at the vertex $v$,
\begin{eqnarray}
\overrightarrow{a}&=&
(a_{e_1},\dots , a_{e_l}, a_{\bar{e}_{l+1}} \rme^{-\rmi k L_{e_{l+1}}},  \dots, a_{\bar{e}_{d}} \rme^{-\rmi k L_{e_{d}}} )^T \ ,\\
\overleftarrow{a}&=&
(a_{\bar{e}_1},\dots , a_{\bar{e}_l}, a_{e_{l+1}} \rme^{\rmi k L_{e_{l+1}}},  \dots, a_{e_{d}} \rme^{\rmi k L_{e_{d}}} )^T \ ,
\end{eqnarray}
Then substituting in the vertex condition (\ref{eq:vertex condition}),
\begin{equation}\label{eq:1}
\overrightarrow{a}=-\big(\mathbb{A}_v+\rmi k \mathbb{B}_v\big)^{-1}\big(\mathbb{A}_v-\rmi k\mathbb{B}_v\big) \overleftarrow{a} \ .
\end{equation}
When the vertex conditions define a self-adjoint Laplace operator the inverse matrix in (\ref{eq:1}) exists and the \emph{vertex scattering matrix},
\begin{equation}\label{eq:vertex scattering}
\sigma^{(v)} = -\big(\mathbb{A}_v+\rmi k \mathbb{B}_v\big)^{-1}\big(\mathbb{A}_v-\rmi k\mathbb{B}_v\big) 
\end{equation}
is unitary \cite{KS99}.

To find the eigenvalues of the Hamiltonian \cite{KS97,KotS99} the vertex scattering matrices are collected into a \emph{graph scattering matrix},
\begin{equation}\label{eq:graph scattering}
\Sigma_{e\, e'} (k) =  \delta_{t(e'),o(e)}  \, \sigma_{e,e'}^{(o(e))} \ ,
\end{equation}
where we regard the two coefficient labels $e$ and $\bar{e}$ for the edge $e$ as directed edges with the origin of $e$ the vertex at $x_e=0$, denoted $o(e)=t(\bar{e})$, and the terminus of $e$ the vertex at $x_e=L_e$, denoted $t(e)=o(\bar{e})$.  
Then a vector of coefficients $\mathbf{a}=(a_1,\dots a_E,a_{\bar{1}}, \dots , a_{\bar{E}})$ defines an eigenfunction on the graph if it is invariant under the \emph{unitary graph evolution operator} $U(k)=D(k) \Sigma(k)$.  Where
\begin{equation}\label{eq:diag phases}
D(k)=\textrm{diag} \{ \rme^{\rmi k L_1},\dots, \rme^{\rmi k L_E} , \rme^{\rmi k L_1},\dots, \rme^{\rmi k L_E} \} \ ,
\end{equation}
a diagonal matrix of the phases acquired by plane waves traversing the $2E$ directed edges.
So for $k\neq 0$ we see $k^2$ is an eigenvalue of the  Hamiltonian if and only if
\begin{equation}\label{eq:secular}
\det \big(\RMI - D(k) \Sigma(k)\big) = 0 \ ,
\end{equation}
which is the \emph{secular equation} of the graph \cite{KS97,KotS99}.

The second approach to graph quantization is now straightforward to describe.  Rather than fixing an operator on the metric graph we can simply consider wave scattering on the metric graph by specifying a set of unitary vertex scattering matrices $\sigma^{(v)}$ for $v \in V$.  Such scattering matrices will not depend on $k$ but we can collect them into a graph scattering matrix $\Sigma$ according to (\ref{eq:graph scattering}) and find values of the wave number $k$ such that $\mathbf{a}$ is invariant under the graph evolution operator which again correspond to solutions of the secular equation (\ref{eq:secular}).  The ensemble of matrices parameterized by $k$ is a unitary stochastic matrix ensemble $U(k)$.  This is the approach introduced in \cite{CC88,CS90} and where spectral properties were first investigated in \cite{T01}.  The advantage of such an approach is that we may be free to choose any unitary vertex scattering matrices, a freedom which can be used to simplify the analysis.

As was pointed out in \cite{KN05} the multiplicity of $k=0$ as a root of the secular equation (\ref{eq:secular}) is not necessarily the multiplicity of zero as an eigenvalue of the Laplacian.  See \cite{BES16} for the characterization of zero modes of the Laplacian.   While in  the case of the wave-scattering quantization one has already assumed $k\neq 0$ apriori.  Consequently, we cannot compare the two versions of graph quantization in the case of zero modes.

\section{Relations Between the Quantizations}\label{sec:lit}

A superficial similarity between the approaches to graph quantization is that they can both be reduced to specifying a $d_v\times d_v$ unitary matrix at each vertex.   The two descriptions are not equivalent however.  From (\ref{eq:vertex scattering}) with vertex conditions defined by the choice of $U_v$ (\ref{eq:alt vertex conditions}),
\begin{equation}\label{eq:vertex scattering U}
\sigma^{(v)} = \big((k-1)\RMI +(k+1)U_v\big)^{-1}\big((k+1)\RMI +(k-1) U_v\big) 
\end{equation}
which, while unitary, is typically dependent on the wave number $k$.

There are many formulations of conditions under which $\sigma^{(v)}$ is independent of $k$ \cite{BerKuch,FKW07,KPS07,KS99}.  The following proposition collects conditions related to the vertex scattering matrix or the matrix pair  $\mathbb{A}_v$ and $\mathbb{B}_v$.  
\begin{proposition}\label{prop:k-independence}  At a vertex $v$ the following are equivalent.
	\begin{arabiclist}
		\item The scattering matrix $\sigma^{(v)}(k)$ is independent of $k$ .
		\item $\mathbb{A}_v \mathbb{B}_v^\dag =0$.
		\item There exists $k\neq 0$ such that $\big(\sigma^{(v)}(k) \big)^2=\RMI$.
		\item $\big(\sigma^{(v)}(k) \big)^2=\RMI$ for all $k$.
	\end{arabiclist}
\end{proposition}
For example, the condition $\mathbb{A}_v \mathbb{B}_v^\dag =0$ is satisfied for standard (or Neumann like) vertex conditions where the wavefunction is continuous at the vertex and the outgoing derivatives sum to zero.  Such vertex conditions can be encoded in matrices,
\begin{equation}\label{eq:standard conditions}
\mathbb{A}_v= 
\begin{pmatrix}
1&-1&0&0&\dots \\
0&1&-1&0&\dots\\
&&\ddots & \ddots & \\
0&\dots &0&1&-1\\
0&\dots &0&0&0
\end{pmatrix}
\qquad  \qquad \mathbb{B}_v= 
\begin{pmatrix}
0&\dots&0\\
\vdots & & \vdots \\
0& \dots &0 \\
1&\dots &1
\end{pmatrix}
\ .
\end{equation}
The energy independent scattering matrix generated by the standard vertex conditions is,
\begin{equation}\label{eq:standard scattering}
\sigma^{(v)}= \frac{2}{d_v} \RMJ - \RMI \ ,
\end{equation}
where $\RMJ$ is a matrix where all the entries are $1$.

However, while there exist energy independent scattering matrices for the Laplace (and similarly Schr\"odinger) operator, most unitary matrices do not square to the identity.  
So, while all $d_v\times d_v$ unitary matrices correspond to unitary vertex scattering matrices via equation (\ref{eq:vertex scattering U}), most $k$-independent unitary matrices are not themselves vertex scattering matrices of a graph Laplacian.  
This includes, for example, the DFT matrix (\ref{DFTmatrix}).  Consequently, it is natural to consider the wave propagation approach to construct a quantum graph as a distinct formulation of a quantum graph.  

In the final section we will see that the two constructions of a quantum graph are not as different as they currently appear.  This is not a new observation \cite{B08} but the situation we present here simplifies previous constructions and has the advantage of being a physical Hamiltonian whose essential properties are easy to describe.  

Before we do this it is worth mentioning a different perspective on the energy dependence of the scattering matrix provided by Cheon, Exner and Turek  \cite{CET10b}.  Using the unique $ST$-form \cite{CET10} of the vertex conditions (\ref{eq:vertex condition}),
\begin{equation}
\left( \begin{array}{cc}
\RMI_{r_\mathbb{B}} & T\\
0&0 \\
\end{array}\right) \mathbf{F}(v)
= \left( \begin{array}{cc}
S & 0\\
-T^* &\RMI_{d-r_\mathbb{B}}  \\
\end{array}\right)\mathbf{F}'(v)
\end{equation}
where $r_\mathbb{B}=\textrm{rank} \, \mathbb{B}$, they show that, in the high energy limit, the scattering matrix is $k$-independent,
\begin{equation}
\lim_{k\to \infty} \sigma^{(v)}(k) = -\RMI_d +2 \left( \begin{array}{cc}
(\RMI_{r_\mathbb{B}}+TT^*)^{-1} & (\RMI_{r_\mathbb{B}}+TT^*)^{-1} T \\
T^* (\RMI_{r_\mathbb{B}}+TT^*)^{-1} &T^* (\RMI_{r_\mathbb{B}}+TT^*)^{-1} T \\
\end{array}\right) \ .
\end{equation}
The $k$-independence of the scattering matrix in the high energy limit is also a feature of metric graphs quantized with a Dirac operator, as we will see in the next section.

\section{Wave Scattering of the Dirac Operator on a Graph}\label{sec:Dirac construction}

On each interval of $\Gamma$ the time independent Dirac equation reads,
\begin{equation}\label{eq:Dirac eqn}
-\rmi \hbar c \alpha \frac{\rmd }{\rmd x_e} \mathbf{f}_e(x_e) +mc^2 \beta \mathbf{f}_e(x_e) = k \mathbf{f}_e(x_e) 
\end{equation}
where $\alpha$ and $\beta$ satisfy the relations $\alpha^2=\beta^2=\RMI$ and $\alpha\beta+\beta\alpha=0$, the Dirac algebra in one dimension.
 If one looks for a faithful irreducible representation of the Dirac algebra, $\alpha$ and $\beta$ can be chosen to be $2\times 2$ Hermitian matrices.  However, in this case time reversal symmetry is not possible unless edges of the original graph $G$ are replaced by pairs of edges of the same length one running in each direction \cite{BH03}.  If one instead wants to keep the physical interpretation of spin deriving from $3+1$ dimensional spacetime symmetries we can regard the equation as the restriction to one dimension of the Dirac equation in three dimensions.  In which case $\alpha$ and $\beta$ are $4\times 4$ matrices which form a reducible representation of the Dirac algebra.  For example, one can take
\begin{equation}\label{eq:alpha beta}
\alpha = \begin{pmatrix}
0&0&0&-\rmi\\
0&0&\rmi&0\\
0&-\rmi &0&0\\
\rmi & 0&0&0
\end{pmatrix} \qquad \beta =
\begin{pmatrix}
1&0&0&0\\
0&1&0&0\\
0&0&-1&0\\
0&0&0&-1
\end{pmatrix} \ .
\end{equation}
A Dirac operator with two component spinors on pairs of oppositely oriented intervals is mathematically equivalent to the Dirac operator with four component spinors.  As we are looking for a physical metric graph operator we will adopt the graph Dirac operator model with four component spinors hence the domain of the Dirac operator will be a subspace of $\bigoplus_{e\in \mathcal{E}} W^{1,2} ([0,L_e])\otimes \mathbb{C}^4$, the Sobolev space of spinors on the interval that are in $L^2$ along with their weak first derivative.

To specify vertex conditions at a vertex $v$ of degree $d$ we define a vector of the first and second components of the spinors at the ends of the edges connecting at $v$ and similarly a vector of the third and fourth components of the spinors,
\begin{eqnarray}\label{eq:Dirac vertex vectors}
\mathbf{F}^+(v)&=&\big(f^{e_1}_1(0),f^{e_1}_2(0), \dots, f^{e_l}_1(0), f^{e_l}_2(0),\nonumber \\
&&\qquad f^{e_{l+1}}_1(L_{e_{l+1}}),f^{e_{l+1}}_2(L_{e_{l+1}}), \dots , f^{e_d}_1 (L_{e_d}),f^{e_d}_2 (L_{e_d}) \big)^T \ , \\ 
\mathbf{F}^-(v)&=&\big(-f^{e_1}_4(0),f^{e_1}_3(0), \dots, -f^{e_l}_4(0), f^{e_l}_3(0),\nonumber \\
&&\qquad f^{e_{l+1}}_4(L_{e_{l+1}}),-f^{e_{l+1}}_3(L_{e_{l+1}}), \dots , f^{e_d}_4 (L_{e_d}),-f^{e_d}_3 (L_{e_d}) \big)^T \ ,
\end{eqnarray}
where we take the first $l$ of the intervals incident at $v$ to have coordinates running in the outgoing direction and the remaining intervals to be equipped with incoming coordinates.  The choice of coordinate orientations does not effect the spectrum, so one is free to imagine having chosen a vertex $v$ where all the intervals meeting at $v$ have a coordinate running in the outgoing direction.  However, it is not possible to satisfy this simultaneously at all  vertices of the graph.

Then, in analogy with the Laplace operator, the domain consisting the subspace of $\bigoplus_{e\in \mathcal{E}} W^{1,2} ([0,L_e])\otimes \mathbb{C}^4$ satisfying vertex conditions,
\begin{equation}\label{eq:Dirac vertex conditions}
\mathbb{A}_v \mathbf{F}^+(v) + \mathbb{B}_v \mathbf{F}^-(v)  = \mathbf{0}
\end{equation}
defines a self-adjoint Dirac operator provided $(\mathbb{A}_v,\mathbb{B}_v)$ has maximal rank and $\mathbb{A}_v\mathbb{B}_v^\dagger$ is Hermitian \cite{BH03}.

Eigenspinors of the Dirac operator are again plane waves.  For positive energy,
\begin{eqnarray}
\mathbf{f}_e(x_e) &=& 
a^{e}_\alpha \begin{pmatrix}  
1\\0\\0\\\rmi \gamma(k) 
\end{pmatrix} \rme^{\rmi k x_e}
+a^e_\beta
 \begin{pmatrix}  
0\\1\\-\rmi \gamma(k)\\0 
\end{pmatrix} \rme^{\rmi k x_e} \nonumber \\
&&
\qquad +a^{\bar{e}}_\alpha \begin{pmatrix}  
1\\0\\0\\-\rmi \gamma(k) 
\end{pmatrix} \rme^{-\rmi k x_e}
+a^{\bar{e}}_\beta
\begin{pmatrix}  
0\\1\\\rmi \gamma(k)\\0 
\end{pmatrix} \rme^{-\rmi k x_e} \ ,
\end{eqnarray}
where 
\begin{equation}
\gamma(k)= \frac{E-mc^2}{\hbar c k} \ , \qquad
E= \sqrt{(\hbar ck)^2+m^2c^4} \ .
\end{equation}
In the case of zero mass $\gamma(k)=1$, and if $m\neq 0$ taking $k$ to infinity $\gamma(k)$ approaches one .  Defining vectors of the coefficients of the incoming and outgoing plane waves at a vertex $v$,
\begin{eqnarray}
\overrightarrow{a}&=&
\big(a_\alpha^{e_1},a_\beta^{e_1}, \dots , a_\alpha^{e_l},a_\beta^{e_l} , \nonumber \\
&&\qquad a^{\bar{e}_{l+1}}_\alpha \rme^{-\rmi k L_{e_{l+1}}}, a^{\bar{e}_{l+1}}_\beta \rme^{-\rmi k L_{e_{l+1}}}, \dots, a^{\bar{e}_{d}}_\alpha \rme^{-\rmi k L_{e_{d}}},a^{\bar{e}_{d}}_\beta \rme^{-\rmi k L_{e_{d}}} \big)^T \ ,\\
\overleftarrow{a}&=& 
\big(a_\alpha^{\bar{e}_1},a_\beta^{\bar{e}_1}, \dots , a_\alpha^{\bar{e}_l},a_\beta^{\bar{e}_l} , \nonumber \\
&&\qquad a^{{e}_{l+1}}_\alpha \rme^{\rmi k L_{e_{l+1}}}, a^{{e}_{l+1}}_\beta \rme^{\rmi k L_{e_{l+1}}}, \dots, a^{{e}_{d}}_\alpha \rme^{\rmi k L_{e_{d}}},a^{{e}_{d}}_\beta \rme^{\rmi k L_{e_{d}}} \big)^T \ ,
\end{eqnarray}
and substituting in the vertex conditions (\ref{eq:Dirac vertex conditions}) the vertex scattering matrix is,
\begin{equation}\label{eq:Dirac vertex scattering}
\sigma^{(v)} = -\big(\mathbb{A}_v-\rmi \gamma(k) \mathbb{B}_v\big)^{-1}\big(\mathbb{A}_v+\rmi\gamma (k)\mathbb{B}_v\big) \ .
\end{equation}
where $\overrightarrow{a} = \sigma^{(v)} \overleftarrow{a}$.
For matrices $\mathbb{A}_v, \mathbb{B}_v$ defining vertex conditions of a self-adjoint Dirac operator, $\sigma^{(v)}$ is unitary \cite{BH03}.
In the massless case $\gamma(k)=1$ and hence the scattering matrix is independent of $k$.   For non-zero mass the scattering matrix is also asymptotically constant for large $k$.  

The absence of energy dependence in the vertex scattering matrices makes the Dirac operator on a metric graph a natural setting to establish a correspondence between a self-adjoint Hamiltonian and the wave-scattering construction \cite{B08}.  
Consider the zero mass Dirac operator, and at each vertex let 
\begin{equation}\label{eq:vertex model}
\mathbb{A}_v=\frac{1}{2} (\RMI-U_v) \qquad \mathbb{B}_v=\frac{\rmi}{2} 
(\RMI +U_v)
\end{equation}
with $U_v$ a $2d_v\times 2d_v$ unitary matrix.
This defines vertex conditions of a self-adjoint Dirac operator with
$\sigma^{(v)}=U_v$ according to (\ref{eq:Dirac vertex scattering}).  So there exist vertex conditions that produce any chosen $k$-independent scattering matrices in the operator model at the cost of allowing two incoming and two outgoing spinor valued plane waves on each edge. 

However, if we set the scattering matrix $U_v= \widehat{U}_v\otimes \RMI_2$ with $\widehat{U}_v$ a $d_v\times d_v$ unitary matrix, the secular equation factorizes, as restricting $U_v$ to this form does not allow spin rotation at vertex scattering events and the Dirac operator decomposes into two operators acting on two component spinors.  The secular equation becomes,
\begin{eqnarray}\label{eq:secular dirac factor}
\det \big(\RMI - (D(k) \widehat{\Sigma} ) \otimes \RMI_2 \big) = 0 \ .
\end{eqnarray}
which reduces to,
\begin{eqnarray}\label{eq:secular dirac factor reduced}
\det \big(\RMI - (D(k) \widehat{\Sigma} ) \big) = 0 \ ,
\end{eqnarray}
where $\widehat{\Sigma}$ is the $k$-independent bond scattering matrix generated by $d_v\times d_v$ vertex scattering matrices $\widehat{U}_v$ according to (\ref{eq:graph scattering}) and $D(k)$ is the diagonal matrix of phases (\ref{eq:diag phases}).  Each solution of the secular equation is an eigenvalue of the Dirac operator with multiplicity two.  Note that this would become multiplicity four due to Kramer's degeneracy if the boundary conditions are time-reversal symmetric which requires $\widehat{\Sigma}= \widehat{\Sigma}^T$, see \cite{BH03}.  

We observe that the secular equation of a Dirac operator on a metric graph, describing a particle with zero mass and where the vertex conditions do not allow spin rotations at the vertices, matches the energy independent secular equation of wave propagation on a quantum graph obtained by specifying unitary vertex scattering matrices directly.  For massive particles this agreement is still obtained asymptotically in the high energy limit.  In this sense the well studied wave-propagation quantization produces a subset of those quantum graphs corresponding to a self-adjoint Hamiltonian operator on a metric graph.  The construction described here simplifies several similar constructions provided to address the same issue in \cite{B08}.   It has the advantage of being both a physically relevant Hamiltonian, the natural restriction of the three dimensional Dirac operator to a metric graph, and being straightforward to describe with the simple requirements of zero mass and no spin rotation at the vertices.

\section*{Acknowledgments}

JH would like to thank Jens Marklof for raising the subject of this article at the QMATH 15 meeting and Gregory Berkolaiko, and the anonymous referees for helpful suggestions.

\end{document}